\documentclass[traditabstract]{aa} 

\usepackage{graphicx}
\usepackage{txfonts}

\begin{document}

\title{Modeling proper motions beyond the Galactic bulge}

\author{Maura Brunetti and Daniel Pfenniger}
\authorrunning{M. Brunetti and D. Pfenniger}

\institute{Geneva Observatory, University of Geneva, CH-1290 Sauverny, 
Switzerland}

\date{}

\abstract{We analyse the radial and tangential velocity fields in the
  Galaxy as seen from the Sun by using as a first approximation a
  simple axisymmetric model, which we then compare with the
  corresponding fields in a barred $N$-body model of the Milky
  Way. This provides a global description of these quantities missing
  in the literature, showing where they take large values susceptible
  to be used in future observations even for sources well beyond the
  Galactic center.  Absolute largest proper motions occur at a
  distance slightly behind the Galactic Center, which are there 1.5
  times larger than the highest local proper motions due to the
  Galactic differential rotation. Large proper motions well beyond the
  Galactic center are well within the current astrometric accuracy.}


\maketitle

\section{Introduction}\label{introduction}

In view of the ongoing and forthcoming advances of astrometry applied
to the Milky Way well beyond the Solar neighborhood, such as the
measurements of the trigonometric parallaxes and proper motions of
distant stellar masers (e.g., Reid et al.~\cite{reid}), or the
upcoming GAIA survey, we develop here a simple first-order model of a
galactic disk making explicit where large proper motions and radial
velocities due to differential Galactic rotation can be expected in
the Galaxy.  It is indeed important to develop quantitative insight on
these observables and separate the effects due to differential
rotation from those due to, for example, a bar.

The level of analysis is elementary and could figure in astronomy
textbooks. Despite literature search and discussions with
astrometrists the points raised here do not seem to be commonly
known. The probable reason for the lack of a previous similar
discussion is that observing proper motions has been considered for a
long time as difficult and limited to a few kpc.  Further, dust
extinction has also limited observations in the Galaxy plane in the
optical.  Therefore, a local analysis has been considered as
sufficient (e.g. Mihalas~\cite{mihalas}, Mihalas \&
Binney~\cite{mihalasbinney}, Binney \&
Merrifield~\cite{binneymerrifield}).

In radio astronomy, however, the need to develop a global vision of
the Galaxy has been reached much sooner.  An analysis close to the
present one concerning the Galaxy radial velocity field has been
already done (see e.g. Burton~\cite{burton}), since extinction in
neutral hydrogen is weak and integrated radial velocity field can be
directly measured over most of the Galaxy disk.

The most interesting aspect of the present analysis concerns the
proper motion field at distances well beyond the Solar neighborhood,
which shows a discontinuous behavior just beyond the Galactic center
\textit{allowing to measure easily stellar proper motions there and
  well beyond}.  The reason is easy to understand if we consider that
the disk beyond the Galactic center is moving in first approximation
with a velocity opposite to the Sun velocity.  The linear velocity
difference of about $2 \times 220\,\rm km\, s^{-1}$ largely exceeds
the inversely to distance decreasing proper motions, leading to proper
motions up to about $440 \,\rm km\, s^{-1} / 9\,kpc \approx 10\,mas
\,yr^{-1}$, a well measurable quantity with current techniques.

Of course the circular velocity is not equal to the average azimuthal
velocity, the velocity that can be observed by averaging proper
motions, due to the non-negligible contribution of the velocity
dispersion to radial support. Especially close to the Galactic center
the velocity dispersion increases to values comparable to $220/\sqrt
3\,\rm km\, s^{-1} \approx 130\, km\, s^{-1} $, therefore one should
see, as regions beyond the Galactic bulge are measured at increasing
distances, a progressive change from individual random motions at
about $130 \,\rm km\, s^{-1} / 8\,kpc \approx 3\,mas \,yr^{-1}$
superposed to a systematic proper motion with respect to an inertial
frame due to the Sun motion of about $220 \,\rm km\, s^{-1} / 8\,kpc
\approx 5\,mas \,yr^{-1}$, progressively increasing to about $400
\,\rm km\, s^{-1} / 10\,kpc \approx 8\,mas \,yr^{-1}$ as we explore
regions 2\,kpc beyond the Galactic center.  This insight was already
exposed by Binney (1995) in a little read paper concerned primarily
with the non-axisymmetric motion effects induced by a bar.  He already
showed that large proper motions can be expected near the Galactic
center even in the absence of a bar.  In any case, the systematics
introduced by differential motion in the bar/bulge region must not be
omitted, as pointed out by Zhao et al.~(\cite{zhao}).

We also check that a fully consistent model of the Milky Way including
the effects of a bar and velocity dispersions does not change the main
conclusions reached with the simple model.  For this, we use one of
the Milky Way $N$-body models of Fux~(\cite{fux}) including a bar and
spiral arms to quantify the differences existing between a fully
consistent, self-gravitating, stable and barred model of the Milky
Way, and the simple axisymmetric model.
  
The plan of the paper is as follows.  In Section 2 we give some
definitions. In Section 3 we develop the first order axisymmetric
model with analytic tools and a more elaborated rotation curve model.
In Section 4 we analyze Fux's $N$-body model and compare it with the
axisymmetric ones and with some observational results. In Section 5 we
present our conclusions.

\section{Decomposition of a differential velocity field}

\begin{figure*}
  \centering
   \includegraphics[width=6.04cm]{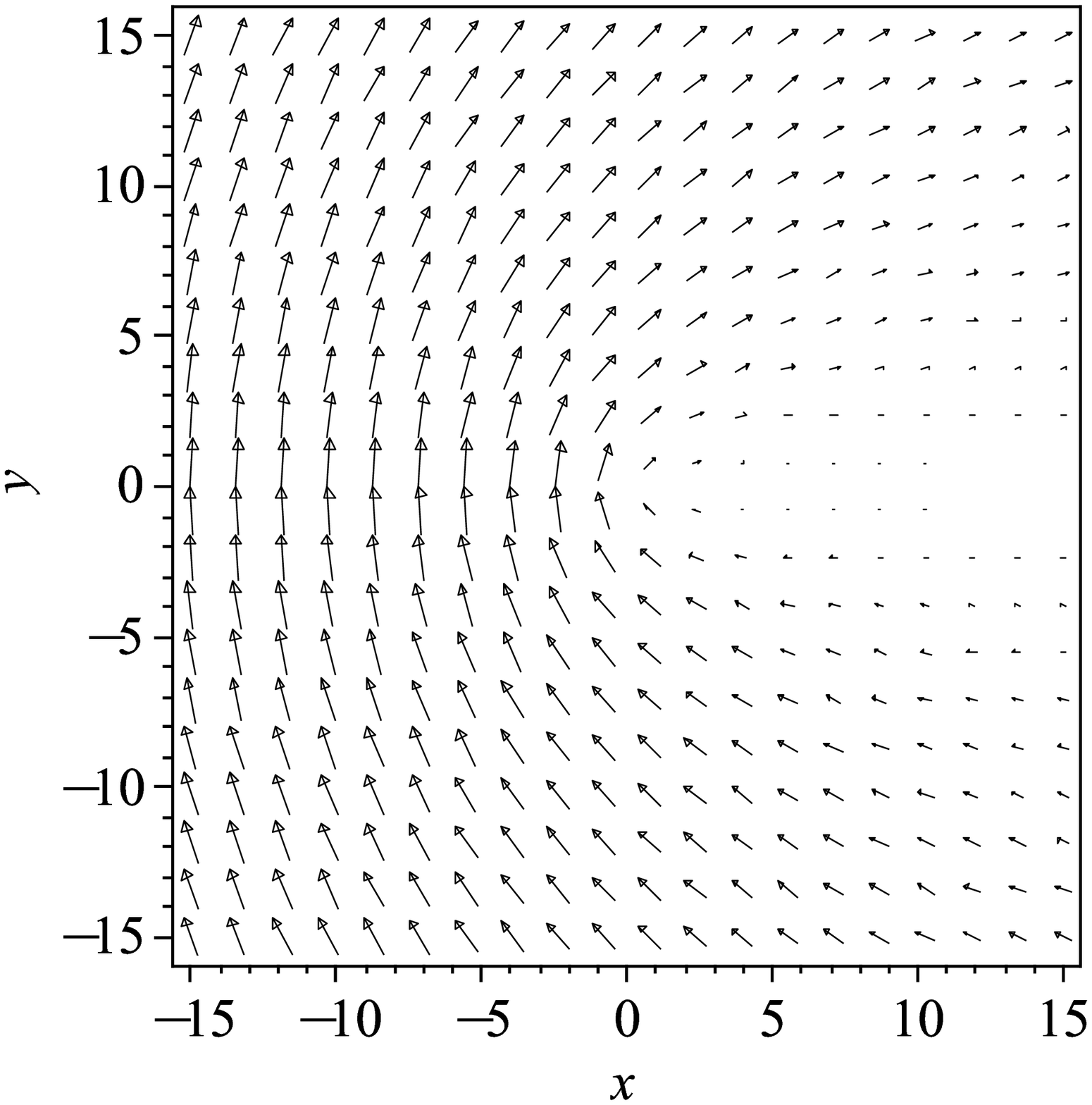}
   \includegraphics[width=6.04cm]{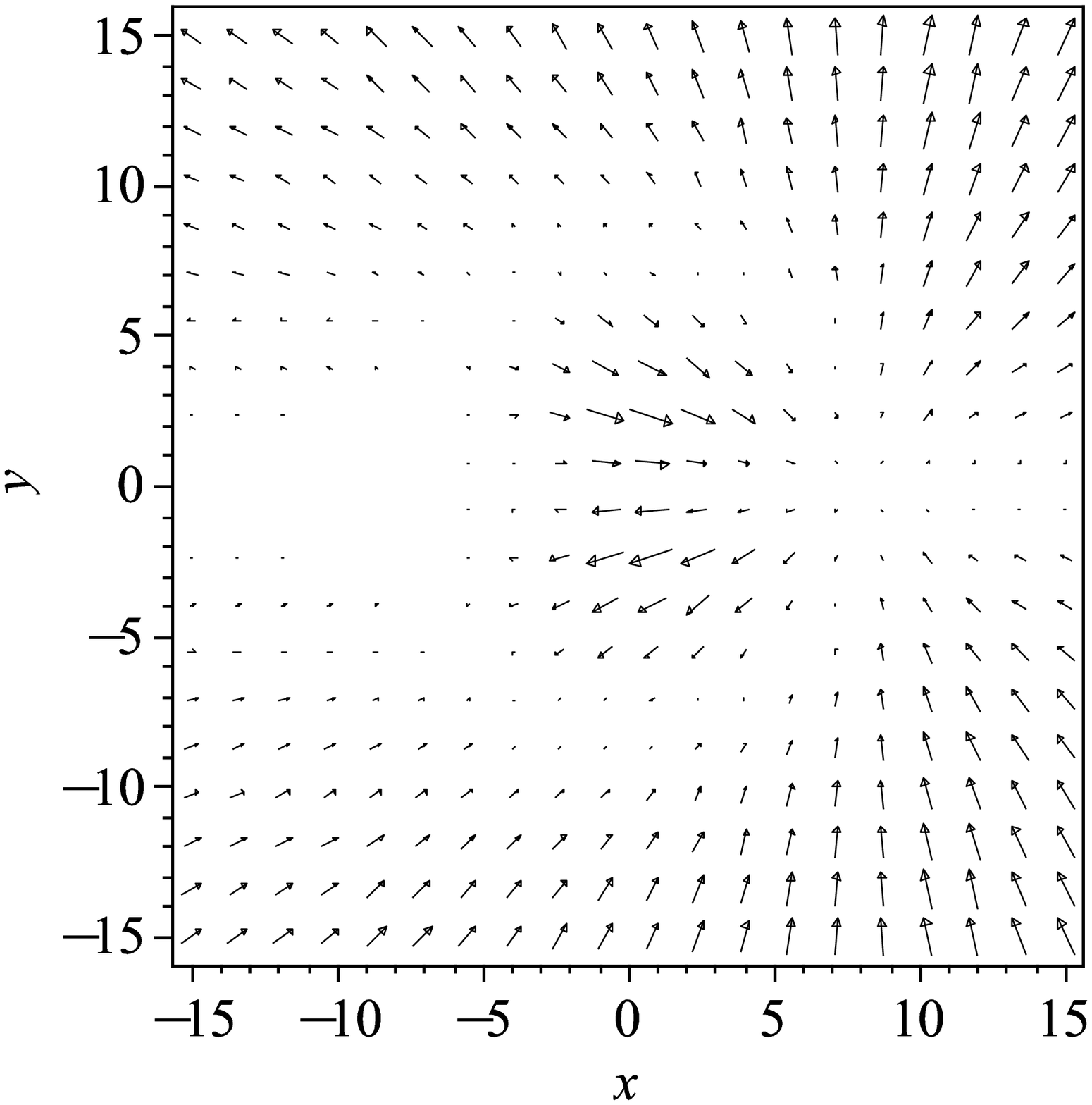}
   \includegraphics[width=6.04cm]{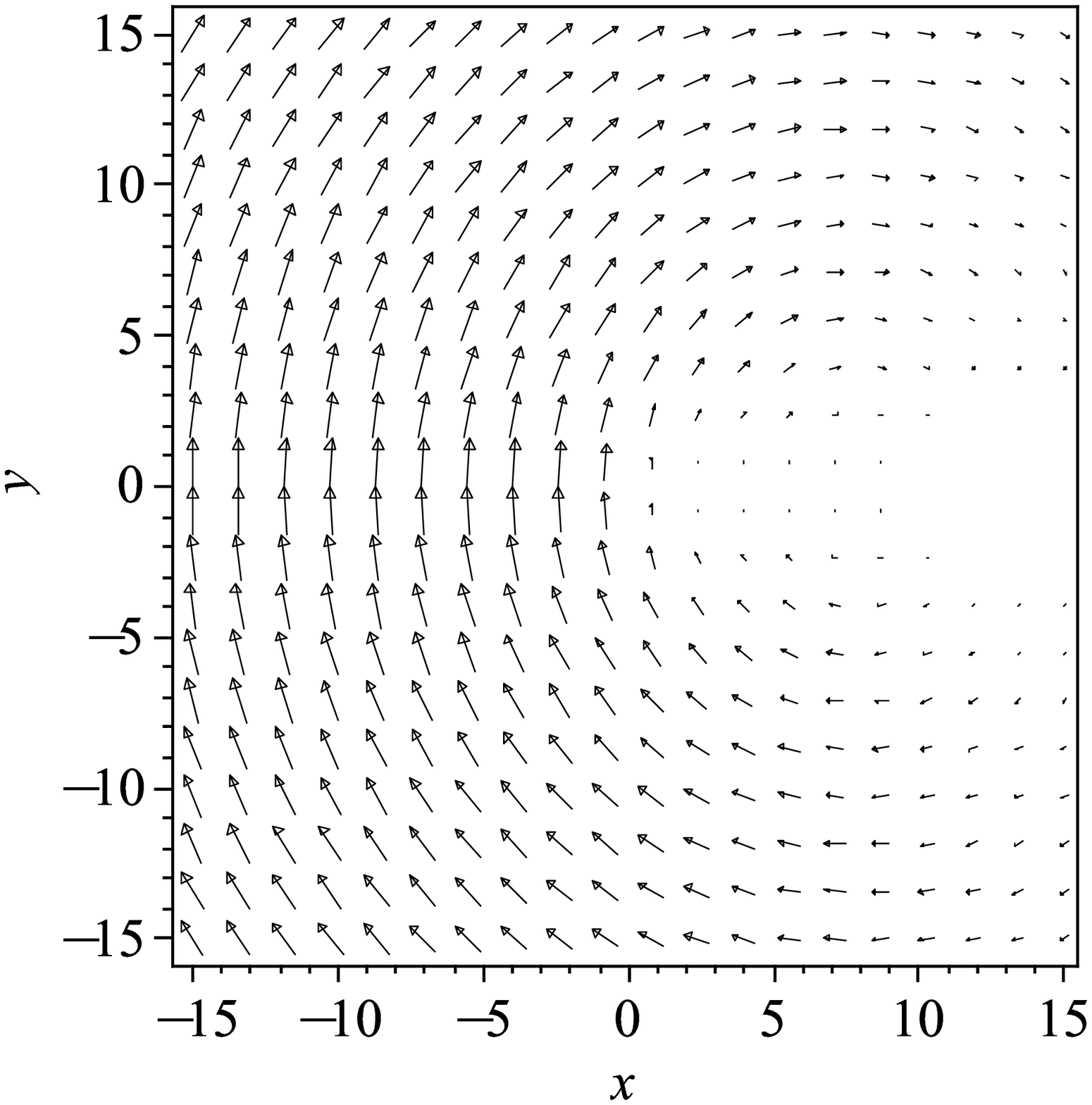}
   \caption{\small The differential velocity field for $h=1$,
   $p=-0.05$, $x_0=8$, $y_0=0$ (left) and its decomposition in radial
   (middle) and tangential (right) components.}
   \label{Fig1}
\end{figure*}

Contrary to the tradition, in the following we express formulas as
much as possible in Cartesian coordinates in order to avoid
trigonometric expressions masking simple geometrical
interpretation. Bold symbols denote vectors.
 
For any velocity field $\mathbf{V}(\mathbf{x}) \in \mathbb{R}^3$,
$\mathbf{x} \in \mathbb{R}^3$ one can define the differential velocity
field relative to the velocity of a particular observer located at
$\mathbf{x_0}$, moving at the velocity of the field
$\mathbf{V}(\mathbf{x_0})$,
\begin{equation}\label{DiffVelField}
  \mathbf{v}(\mathbf{x}) \equiv \mathbf{V}(\mathbf{x})- 
  \mathbf{V}(\mathbf{x}_0). 
\end{equation}
The scalar radial velocity field $v_r(\mathbf{x})$ with respect to the
observer, the usually observable quantity derived from the Doppler
shift, is the projection of $\mathbf{v}(\mathbf{x})$ along the line of
sight unit vector
\begin{equation}\label{UnitVector}
\mathbf{n}_r (\mathbf{x}) \equiv (\mathbf{x} - \mathbf{x}_0)/ d,
\end{equation}
where $d \equiv |\mathbf{x} - \mathbf{x}_0|$ is the distance to the
observer location. The vector velocity field
$\mathbf{v}_r(\mathbf{x})$ is proportional to $v_r$ and directed in
the radial direction $\mathbf{n}_r$.
\begin{equation}
v_r(\mathbf{x})\equiv \mathbf{v}(\mathbf{x}) \cdot \mathbf{n}_r  \, , \qquad
\mathbf{v}_r(\mathbf{x}) \equiv v_r(\mathbf{x})\,  \mathbf{n}_r.
\end{equation}

The tangential velocity field is the difference between the velocity
field and the radial velocity field,
\begin{equation}
\mathbf{v}_t(\mathbf{x}) = \mathbf{v}(\mathbf{x}) - 
\mathbf{v}_r(\mathbf{x}).
\end{equation}

The observable proper motion field $\mathbf{\mu}$ is the tangential
velocity field $\mathbf{v}_t(\mathbf{x})$ divided by the distance $d$
to the observer

\begin{equation}
\mathbf{\mu}  = {\mathbf{v}_t(\mathbf{x}) \over d}.
\end{equation}
This is a 3-dimensional vector field orthogonal to the line-of-sight,
tangential to the celestial sphere. It can be converted to two angular
velocity components for the desired spherical coordinate system using
standard spherical trigonometry formulas. Here we will only use the
magnitude of $\mathbf{\mu}$.

\section{Axisymmetric models}

\begin{figure*}
   \centering
   \includegraphics[width=9cm]{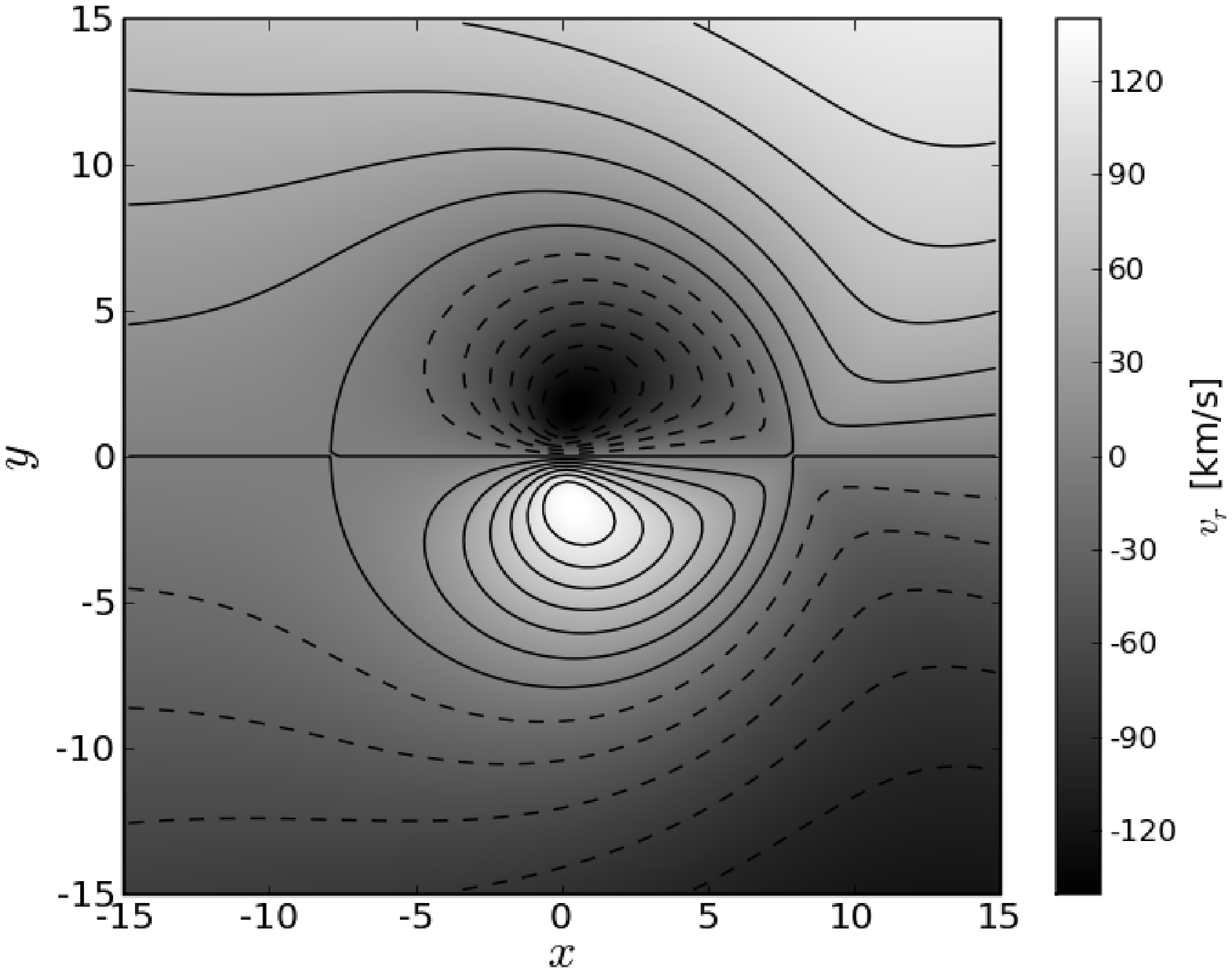}%
   \includegraphics[width=9cm]{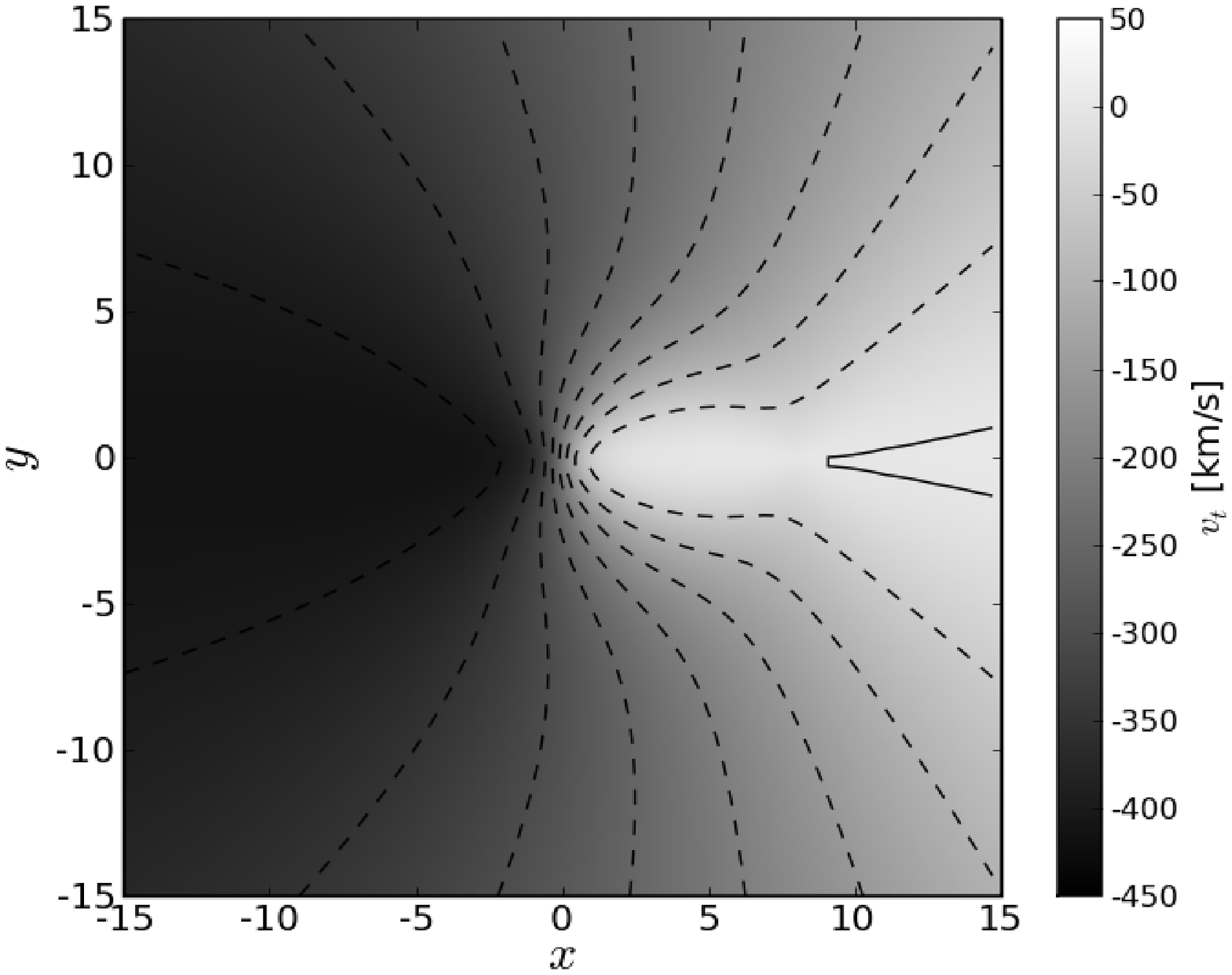}
   \caption{\small Contour and gray map of the magnitude of the
     differential velocity field as in Fig.~\ref{Fig1} decomposed in
     its radial (left) and tangential (right) components.  The
     dark/light grays correspond to low/large values.}
   \label{Fig2}%
\end{figure*}

\begin{figure*}
   \centering
   \includegraphics[width=9cm]{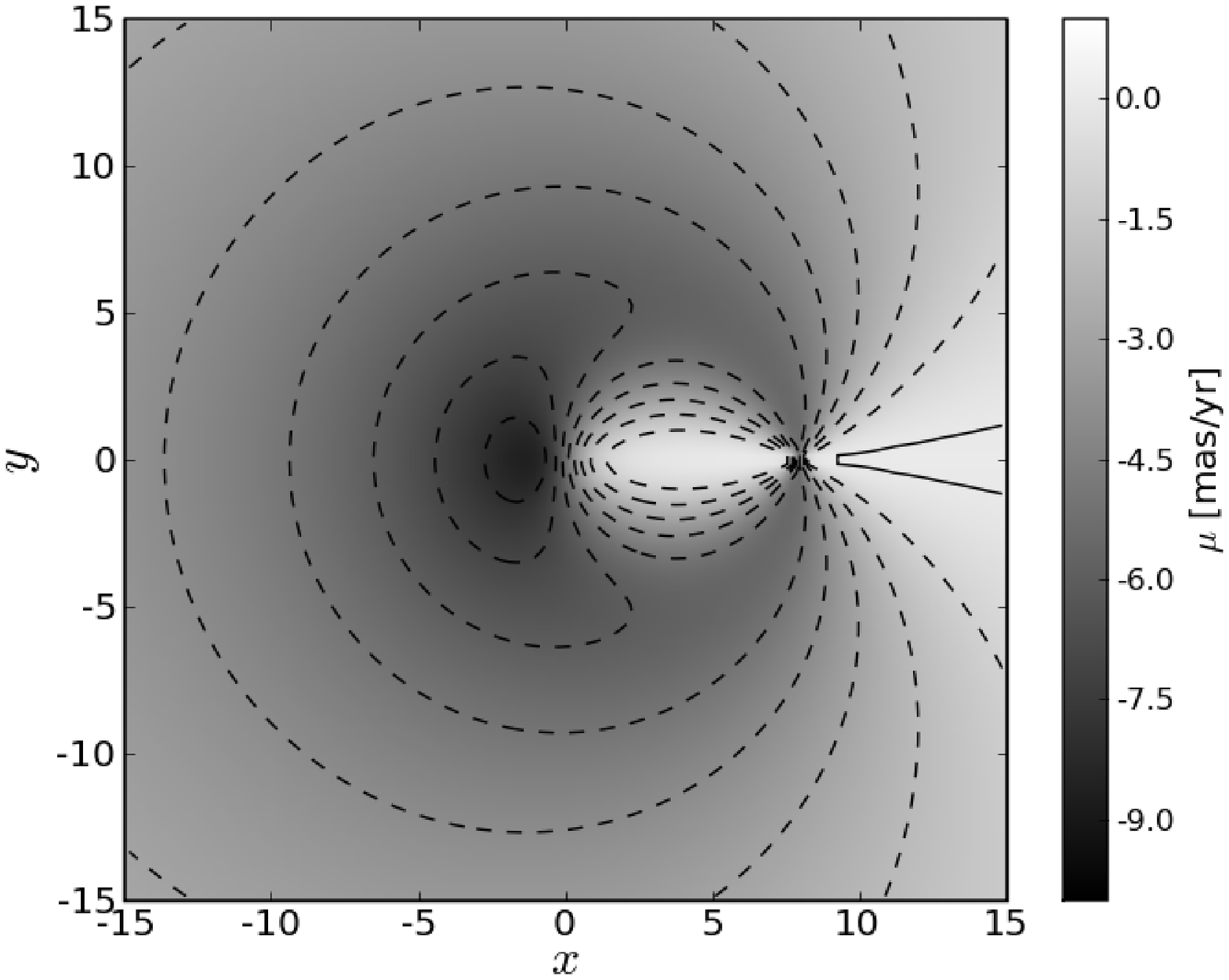}%
   \includegraphics[width=8cm]{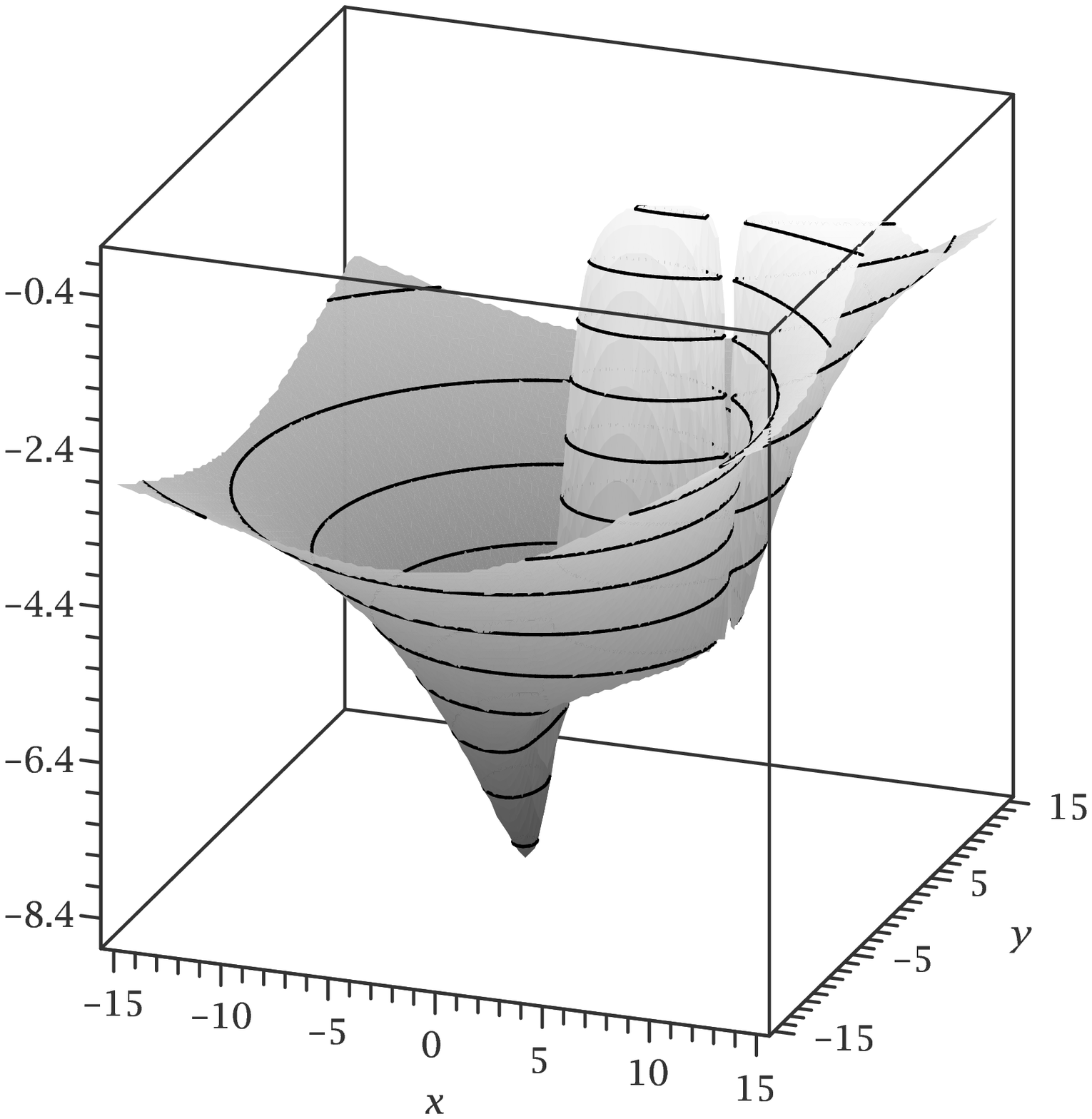}
   \caption{\small On the left, contour map of the magnitude of the
     proper motion $\mu$ as observed from the observer's position at
     $x_0=8, y_0=0$ as in Fig.~\ref{Fig1}.  The dark/light grays
     correspond to low/large values.  On the right, the side view of
     the same contour map shows the relative importance of proper
     motion, with the most negative, largest absolute value slightly
     beyond the Galactic center as seen from the observer's position.}
            \label{Fig3}%
\end{figure*}

The purpose of an axisymmetric model is to provide some insight into
rotation curves resembling the Galaxy one.  Therefore we choose first
a particular representative but simple model of the Milky Way rotation
curve that can be handled analytically.  In a second step we use a
more realistic but still axisymmetric model of the Milky Way
(Kalberla~\cite{kalberla}).

\subsection{Simple model}

We consider a simple extension of the popular logarithmic axisymmetric
galactic potential.  We use cylindrical coordinates $(R,z)$
($R^2=x^2+y^2$, with the origin at the Galactic center $x=y=z=0$).
The potential has the form:
\begin{equation}
\Phi(R,z) \equiv 
   \left\{ 
    \begin{array} { cc}  
       { v_0^2 \over p}
         \left(1+\frac{R^2}{h^2} + \frac{z^2}{z_0^2}\right)^{p/2}, 
          \quad  & \textrm{for}\  p \neq 0 ,\\
        \frac{v_0^2}{2} \ln\left(1+\frac{R^2}{h^2} 
           + \frac{z^2}{z_0^2}\right),      
           \quad & \textrm{for}\  p = 0 , \\
    \end{array}
   \right.
\end{equation}
where $p$ is a parameter, $p \in [-1,2]$, allowing to study different
rotation curves going from an asymptotic Keplerian model ($p=-1$), to
a constant rotation curve ($p \to 0$) (the usual logarithmic
potential), to an harmonic potential ($p=+2$). The constants $h$ and
$z_0$ denote the respective horizontal and vertical scale lengths.

The circular velocity reads
\begin{equation}
v_c = \left( R\partial_R\Phi|_{z=0} \right)^{1/2} =
 v_0  {R \over h}  
\left[ 1 + {R^2 \over h^2} \right]^{p-2\over 4}. 
\end{equation}
When $p=0$ the rotation curve $v_c(R)$ is asymptotically flat (for $R
\gg h$), while for negative/positive values of $p$ it
decreases/increases.  For $p=2$, it reduces to a straight line, a
solid rotation corresponding to a constant density mass model. For
$p=-1$ it decreases asymptotically like a Keplerian model where the
mass is concentrated at the center.  When, additionally to $p=0$, we
send $h\to 0$, the rotation curves becomes a constant $v_c = v_0$ for
all $R>0$, which corresponds to Mestel's disk rotation curve.

We could have chosen to start directly with a rotation curve
expression which represents the effective average azimuthal rotation
field.  However not any rotation field is consistent with a solution
of Jeans' equations and a positive mass distribution.  The advantage
of the circular rotation curve derived from a potential is that it
gives a rotation field consistent with a specific potential-density
pair, while its drawback is that the circular rotation speed
overestimates the effective mean rotation speed.
 
The circular velocity vector field for a clockwise circular rotation
like in the Milky Way reads,
\begin{equation}
\mathbf{V}(\mathbf{x})= -v_c \, \mathbf{n}_z \wedge \mathbf{n}_R =  
  \frac{v_c}{R} \left\{  y , -x , 0 \right\},  
\end{equation}
where $ \mathbf{n}_R \equiv\{x,y,0\}/R $, and $ \mathbf{n}_z
\equiv\{0,0,1\}$ are unit vectors in the galactic plane and
perpendicular to it, respectively.  The differential vector fields
$\mathbf{v}(\mathbf{x})$, $\mathbf{v}_r(\mathbf{x})$, and
$\mathbf{v}_t(\mathbf{x})$, as well as $\mathbf{\mu}(\mathbf{x})$, can
be derived in analytic form using the definitions given above.

The radial vector field reads
\begin{equation}
{v_r(\mathbf{x}) \over v_0} =
 { x y_0 - y x_0 \over hd}  
 \left[ 
   \left( 1 + {R^2 \over h^2} \right)^{p-2 \over 4} -
  \left( 1 + {R_0^2 \over h^2 } \right)^{p-2\over 4} 
 \right] ,
\end{equation}
where $d$ is the distance to the Sun.  This expression vanishes along
the line $xy_0 = yx_0$, and along the circle $R=R_0$.

The tangential vector field reads
\begin{equation}
{v_t(\mathbf{x}) \over v_0} =  
   {x x_0\!+\!y y_0\!-\!R^2 \over hd} \left( 1 \!+\! {R^2 \over h^2} \right)^{p-2\over 4} \!\!+
   {x x_0\!+\!y y_0\!-\!R_0^2  \over hd}  \left( 1 \!+\! {R_0^2 \over h^2} \right)^{p-2\over 4} 
 \!\!\!\!.
\end{equation}
Here the location of vanishing transverse velocity is not simple.
$v_t(\mathbf{x})$ is symmetric with respect to the line $xy_0=yx_0$.
At the limit $d\to 0 $ the field is proportional to $d$, that is,
continuous but not differentiable.  This is a well know result from
the classical textbook analysis of the local motions around the Sun,
where in term of Oort's constants $A
=\frac{1}{2}(v_c/R-dv_c/dR)|_{R_0}$ and $B
=-\frac{1}{2}(v_c/R+dv_c/dR)|_{R_0}$, we have
\begin{equation}
v_t =  d (A\cos(2l) + B).
\end{equation}  
This quantity converges to 0 at $d\to 0$, but with a slope depending
on $l$.
   
\begin{figure*}
   \centering
   \includegraphics[width=8cm]{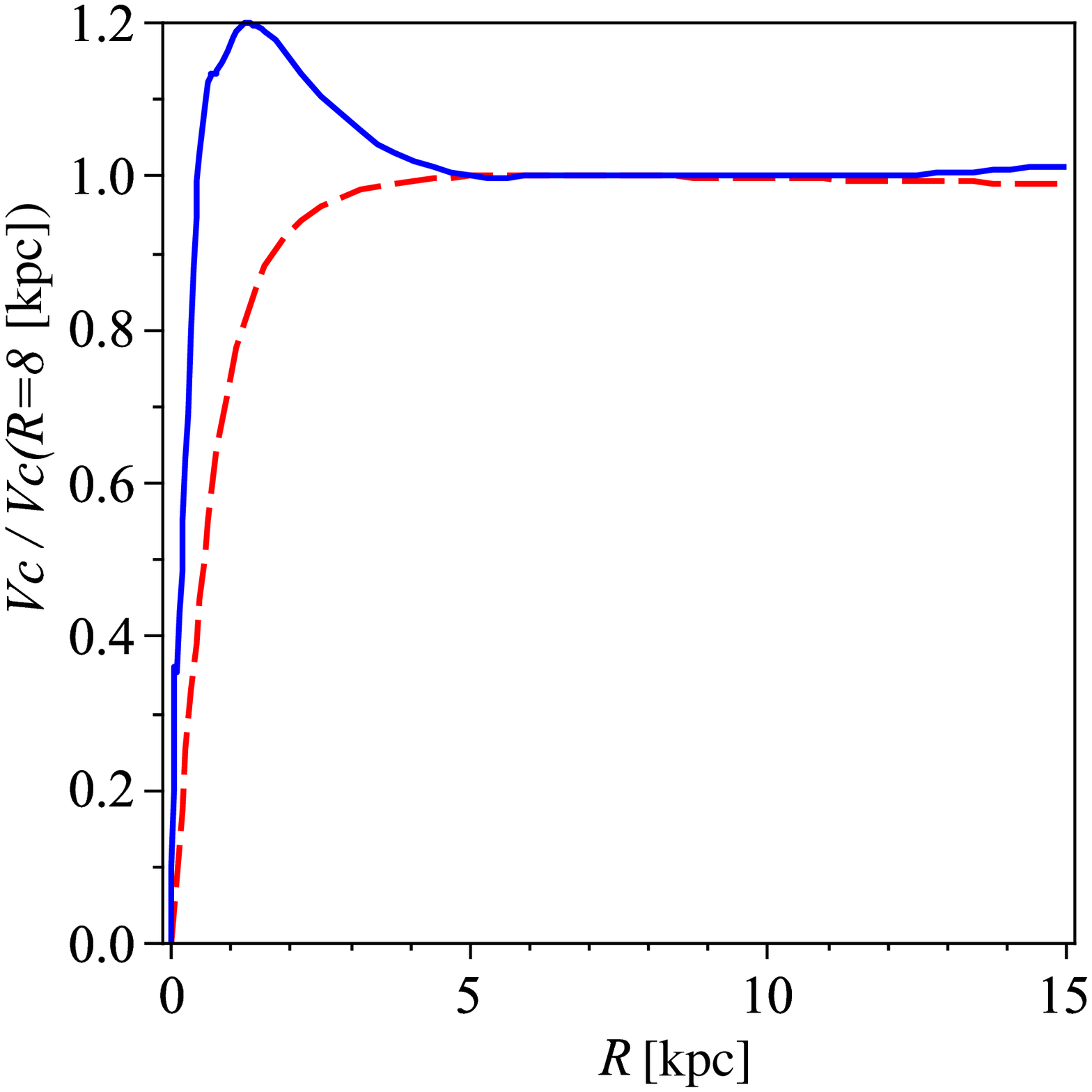}%
   \includegraphics[width=8cm]{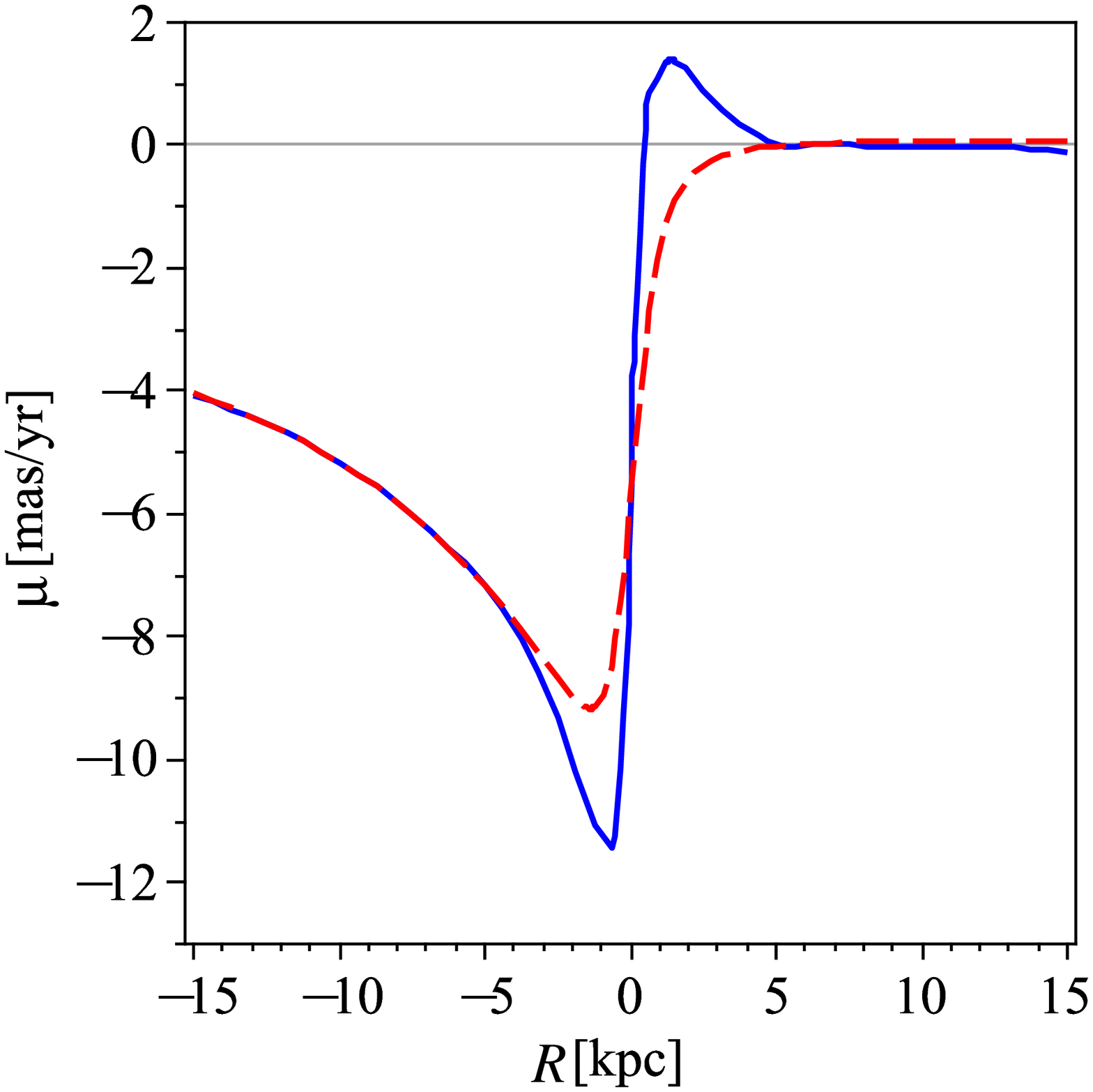}%
   \caption{\small On the left, rotation curve for the analytic model
     (dashed red line, same parameters as in Fig.~\ref{Fig1}) and the
     Kalberla model (solid blue line).  On the right, proper motion
     $\mu_0$ along the line passing through the Galactic center as
     observed from the observer's position at $R_0=8$, in $\rm{mas
       \,yr}^{-1}$ for $v_0 =220~\rm km\, s^{-1}$ and kpc length unit,
     for the analytic model (dashed red line, same parameters as in
     Fig.~\ref{Fig1}) and the Kalberla model (solid blue line).}
   \label{Fig4}
\end{figure*}

For the sake of illustration we show some figures for relevant cases.
We set $h=1$ and $p=-0.05$ to study conditions similar to those
occurring in our Galaxy for a population of objects with a positive
velocity dispersion, which sees its effective rotation decreased by
the asymmetric drift.  We put the observer at the position
$\mathbf{x}_0 = \{8,0,0\}$, which, in kpc unit, corresponds
approximately to the Sun position with respect to the Galactic center.

The differential velocity field is shown in the left panel of
Fig.~\ref{Fig1} and the radial component is shown in the middle panel
of Fig.~\ref{Fig1} and in the left panel of \ref{Fig2}. Along the
$x$-axis, where the observer is placed, the radial velocity vanishes,
as well as along the circular orbit of the observer at radius $R
\simeq 8$, as noticed above.  The maximum radial velocities within the
observer orbit are located on both sides of the Galactic center
direction, and at distances slightly shorter than the Galaxy center.
This field has been used extensively in radio astronomy where the
neutral hydrogen radial velocity can be measured over most of the
Galaxy.

The tangential velocity field is shown in the right panels of
Figs.~\ref{Fig1} and \ref{Fig2}. As can be expected the tangential
velocity is small in the direction of the Galaxy center on the same
side as the observer, because the matter turns in the same direction
at a similar speed there, while the side of the Galaxy beyond the
Galactic center sees large tangential velocities because the matter
turns in opposite directions.  But since only the proper motions are
observable, we must consider the proper motion field, shown in
Figs.~\ref{Fig3}.

There is a strongly negative proper motion at the point $(x= -1.395,
y=0)$ in a region behind the Galactic center, a region characterized
by large amplitude absolute proper motion, $\mu \simeq -8.4\,\rm{mas
  \,yr}^{-1}$ for $v_0 = 220~\rm km\, s^{-1}$ and kpc length unit, 1.5
times larger than the highest local proper motion contribution due to
galactic rotation ($\mu\simeq -5.6\,\rm{mas \,yr}^{-1}$).  For
positive values of the $x$ coordinate, the proper motion $\mu$ is
nearly zero, except at the observer position where a sharp
discontinuity occurs.  This follows from the above remark about the
local tangential differential velocity close to the Sun (Eq.~(11)).
Since $\mu = v_t/d$, then $\lim_{d\to 0}\mu = A\cos(2l)+B$, i.e., is
not single-valued.

The horizontal proper motion as a function of distance along the line
Sun-Galactic center takes the following form,
\begin{equation}
 \mu_0(x) = -{v_0 \over x - x_0} \left[
    {x \over h}\left( 1+ {x^2 \over h^2}\right)^{p-2 \over 4} - 
    {x_0\over h}\left( 1+ {x_0^2\over h^2} \right)^{p-2\over 4}
    \right].
\end{equation}
The zeros and extrema of $\mu_0(x)$ do not have simple analytic
expressions.  This curve is shown in Fig.~\ref{Fig4} (right panel,
dashed red line) for a representative choice of the free
parameters. Clearly the proper motion maximum amplitude occurs for
objects located slightly beyond the Galactic center.

\subsection{Kalberla's model}

Here we check with a more elaborated rotation curve model that our
basic result holds, i.e. that the proper motions slightly beyond the
Galactic center have largest amplitudes.

We take the model of Kalberla~(\cite{kalberla}) and extract 40 data
points from the rotation curve figure. These data are then represented
as a cubic B-spline curve in the symbolic software package Maple 13.
This B-spline object can be considered as a usual function on which
standard operations, including differentiation, can be applied.  The
analysis can be repeated exactly as for the previous model.

Since the results turn out to be qualitatively the same, we only show
the initial rotation curve and the final proper motion curve along the
direction of the Galactic center in Figs.~\ref{Fig4} (solid blue
lines).  The main difference occurs near the center inside the inner 3
kpc where the rotation curve presents a bump and a steeper raise.
This feature may be due to the streaming motions in the bar and, for
applying it to observed stars, it should be corrected from the large
velocity dispersion there by decreasing the average circular motion.
Thus the higher proper motion amplitude (up to 12~$\rm{mas
  \,yr}^{-1}$) obtained here beyond the Galactic center is probably an
upper bound of what could be observed.

The differences between the models are largest precisely in the
bulge-bar region where the absolute proper motions are largest,
i.e., easier to obtain.  Proper motions therefore represent a powerful
observable for discriminating models of the bulge-bar kinematics.

\begin{figure}
   \centering
   \includegraphics[width=9cm]{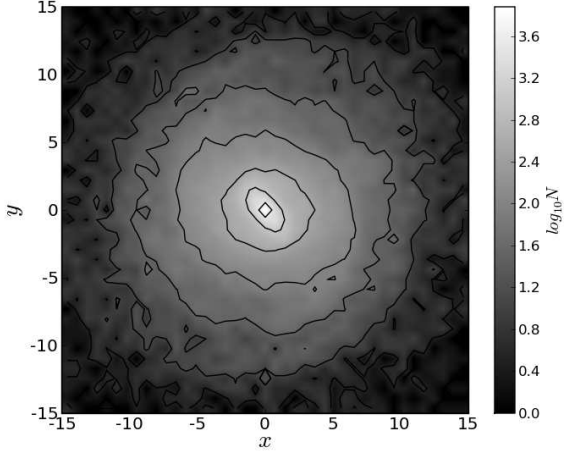}
   \caption{\small Density distribution for the barred $N$-body model
{\tt m08t3200}.}
   \label{Fig5}%
\end{figure}

\begin{figure}
   \centering
   \includegraphics[width=9cm]{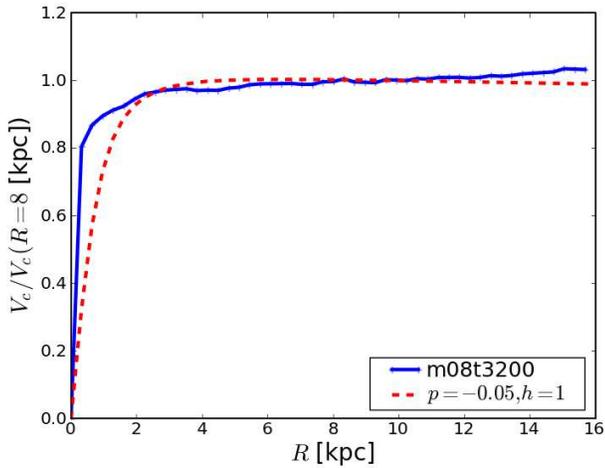}
   \caption{\small Rotation curve for the barred $N$-body model {\tt
       m08t3200} (solid blue line) and the analytic model (dashed red
     line, same parameters as in Fig.~\ref{Fig1}).}
   \label{Fig6}%
\end{figure}

\begin{figure}[t]
   \centering
   \includegraphics[width=9cm]{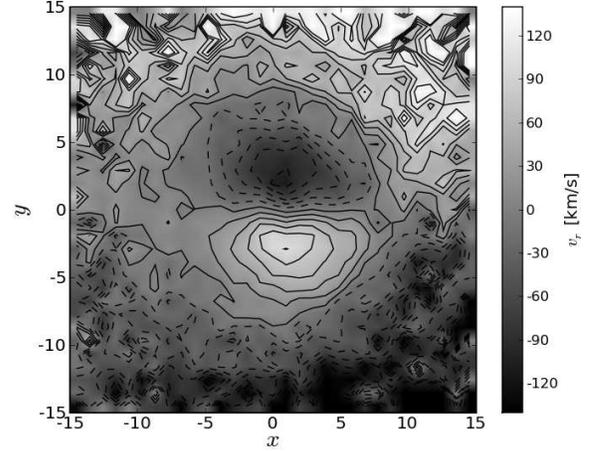}
   \includegraphics[width=9cm]{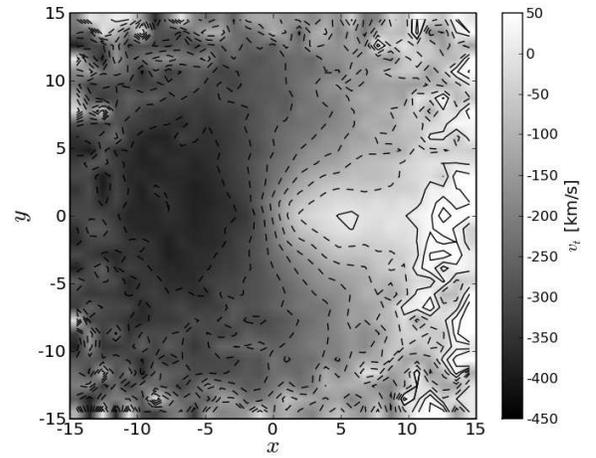}
   \caption{\small Contour map of the magnitude of the differential
     velocity field for the barred $N$-body model {\tt m08t3200}
     decomposed in its radial (top) and tangential (bottom)
     components.}
   \label{Fig7}%
\end{figure}

\section{Barred $N$-body model}

The previous models based on rotation curves do not take into account
the decrease of average rotation due to asymmetric drift, nor the
effects of a bar. Both effects are largest in the bulge-bar
regions, therefore it is useful to compare our previous models to a
fully self-consistent model of the Milky Way including a bar.

Here we analyse one of the barred models of the Milky Way obtained by
Fux~(\cite{fux}) and we compare it with the axisymmetric models
discussed in the previous section and with observational results.

Fux's barred models result from the self-consistent $N$-body evolution
of a bar-unstable axisymmetric model, with $10^5$ particles in the
nucleus-spheroid component (which is well suited to represent the
nuclear bulge and the stellar halo) and in the disk component, and
$10^5$ particles in the dark halo component (which has been added to
ensure a flat rotation curve at large radii).  The spatial location of
the observer has been constrained by the $K$-band map, obtained by the
Diffuse Infrared Background Experiment (DIRBE) on board the COBE
satellite and corrected for extinction by dust, while masses have been
scaled according to the observed radial velocity dispersion of M
giants in Baade's window.  Among the models obtained by imposing such
constraints, the model named {\tt m08t3200} is recommended in
Fux~(\cite{fux}) since it reproduces satisfactorily the kinematics of
disk and halo stars in the Solar neighborhood.

The density distribution of {\tt m08t3200} is shown in
Fig.~\ref{Fig5}.  By a rotation of the particles the observer has been
moved to the position ${\bf x}_0 = \{8,0,0\}$ in kpc unit. The angle
$\phi$ between the line joining the observer to the Galactic center
and the major axis of the bar equals to $\phi = 25^{\circ}$ and the
corotation radius is $R_L = 4.8$~kpc (Fux~\cite{fux}).

The rotation curve is shown in Fig.~\ref{Fig6}, solid blue line.  For
comparison, we also show the rotation curve for the analytic model
discussed in the previous section, for $p = -0.05$, $h=1$. As in
Kalberla's model, the main difference occurs near the central 3~kpc,
with the absence of a bump.

\begin{figure*}
   \centering
   \includegraphics[width=9cm]{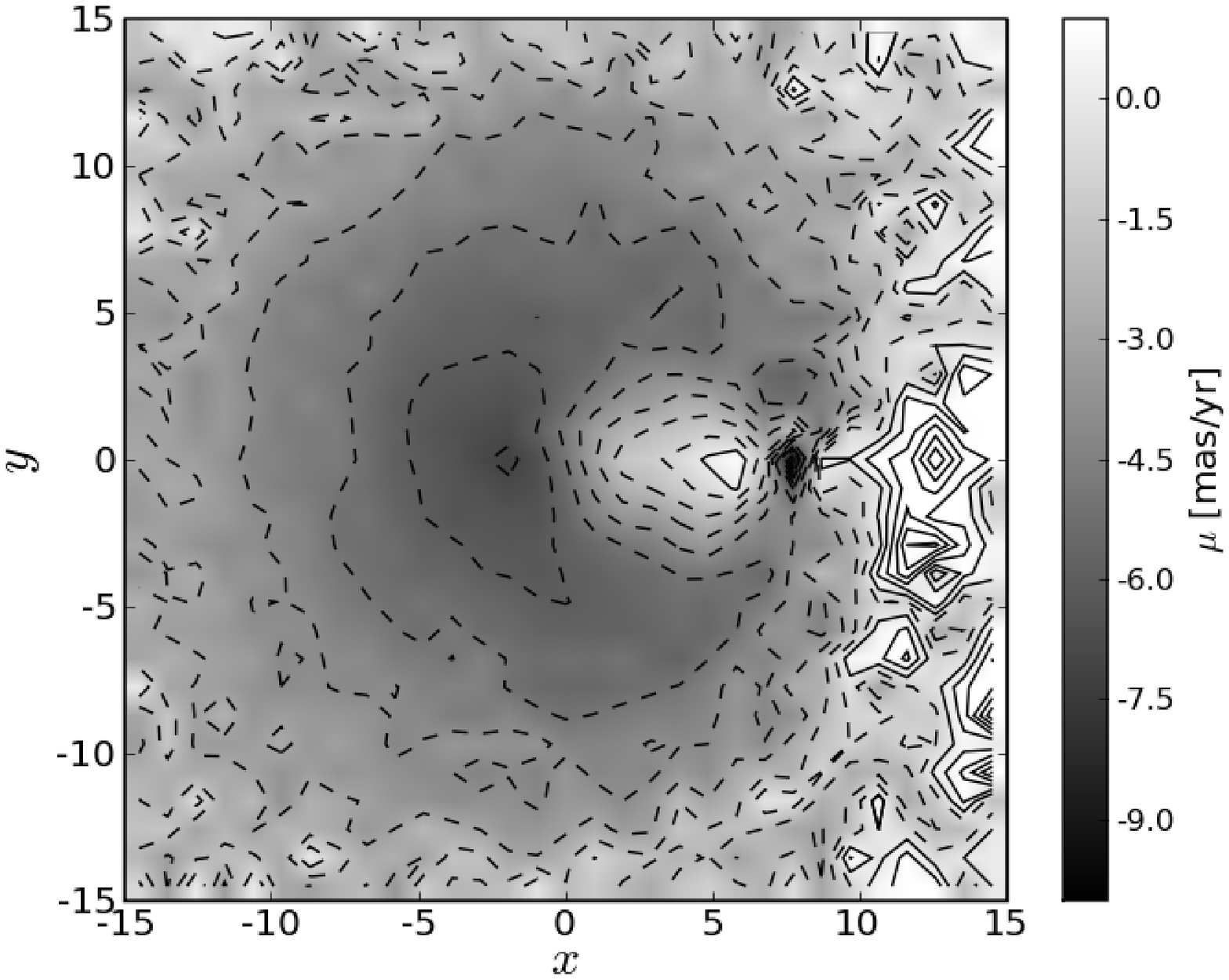}
   \includegraphics[width=9cm]{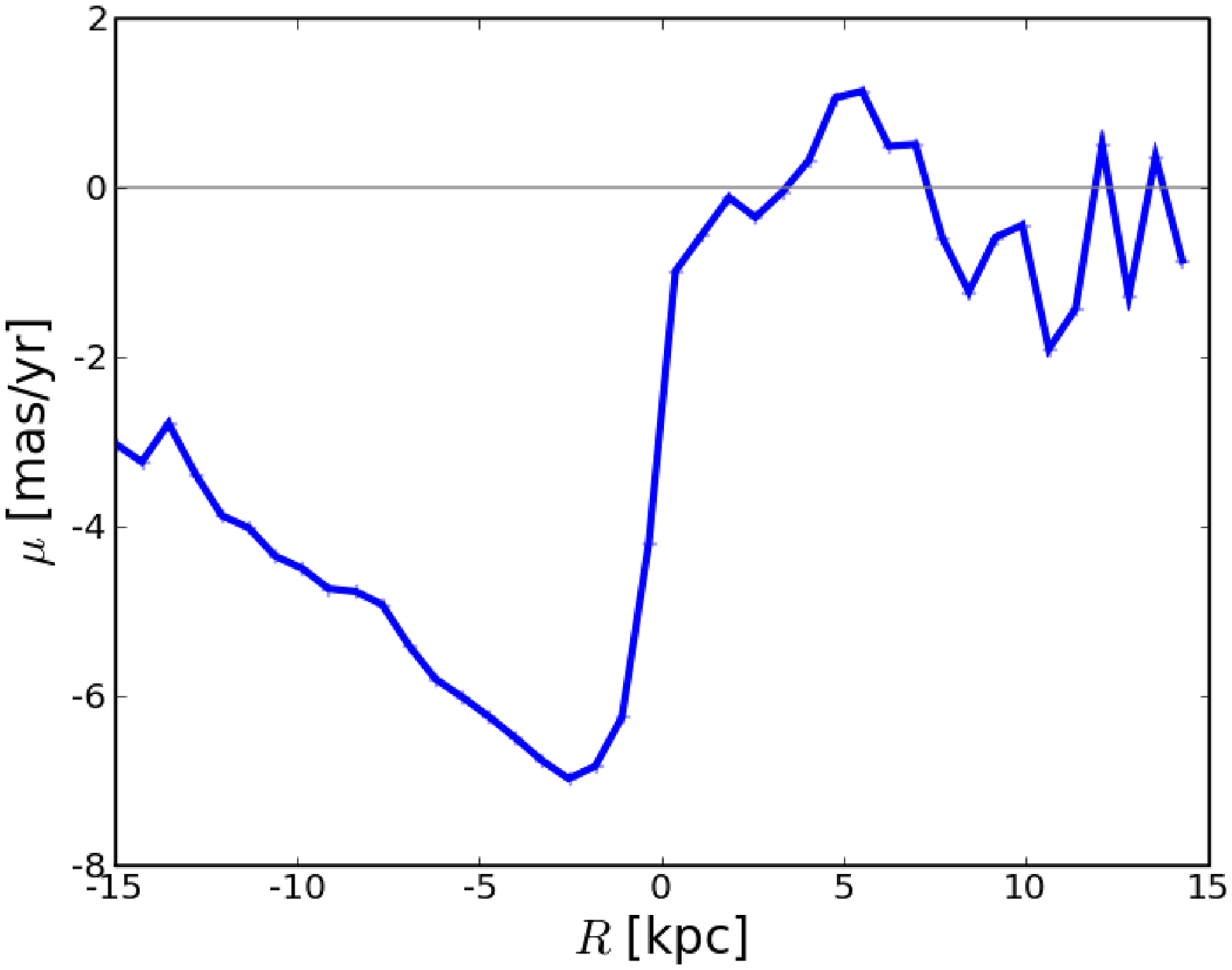}
   \caption{\small Contour map of the magnitude of the proper motion
     $\mu$ as observed from the observer's position at $x_0 = 8$, $y_0
     = 0$ (left) and proper motion $\mu_0$ along the line passing
     through the Galactic center (right) for the barred $N$-body model
     {\tt m08t3200}.}
   \label{Fig8}%
\end{figure*}

\begin{figure*}
   \centering
   \includegraphics[width=6.08cm]{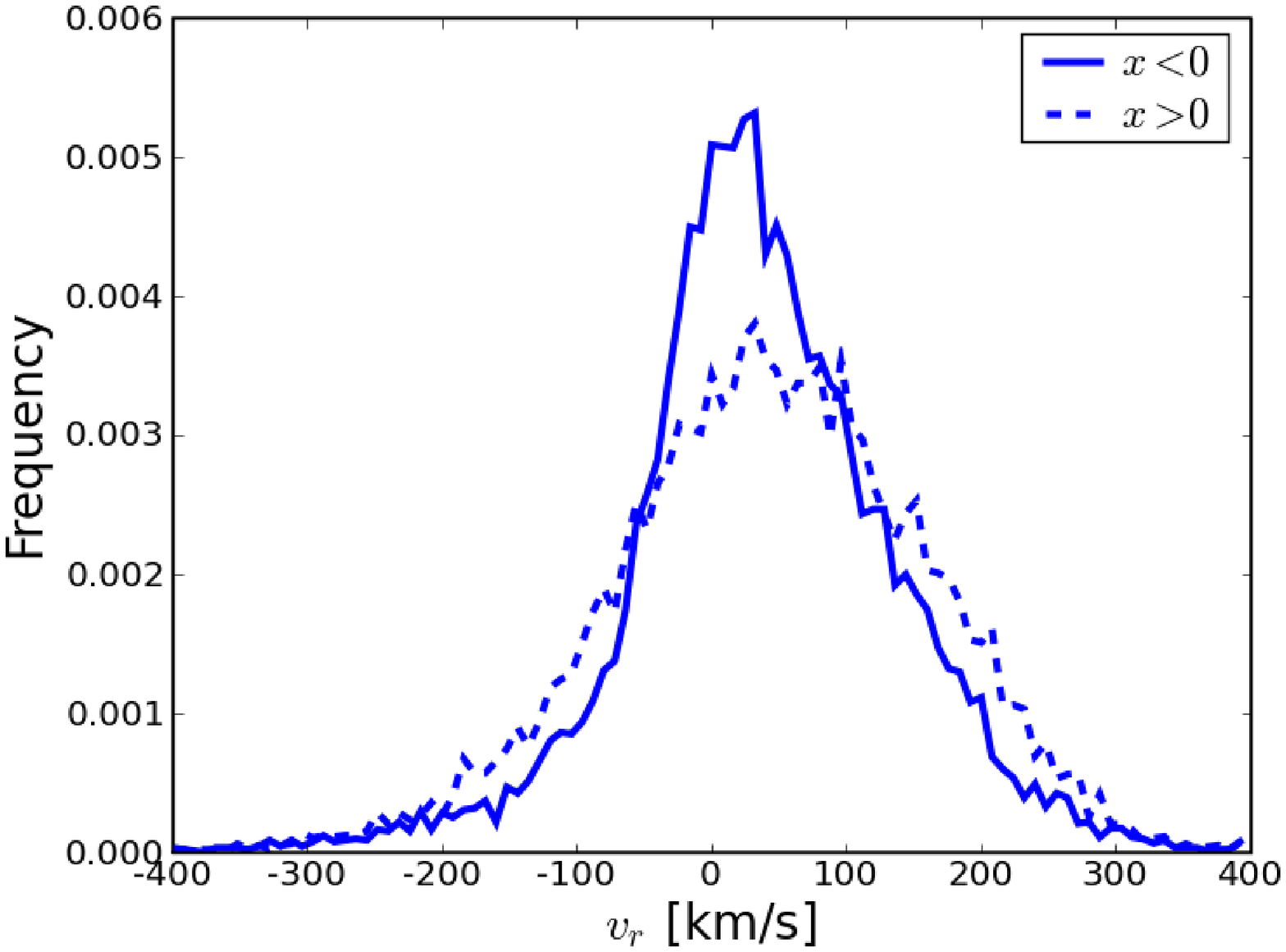}
   \includegraphics[width=6.08cm]{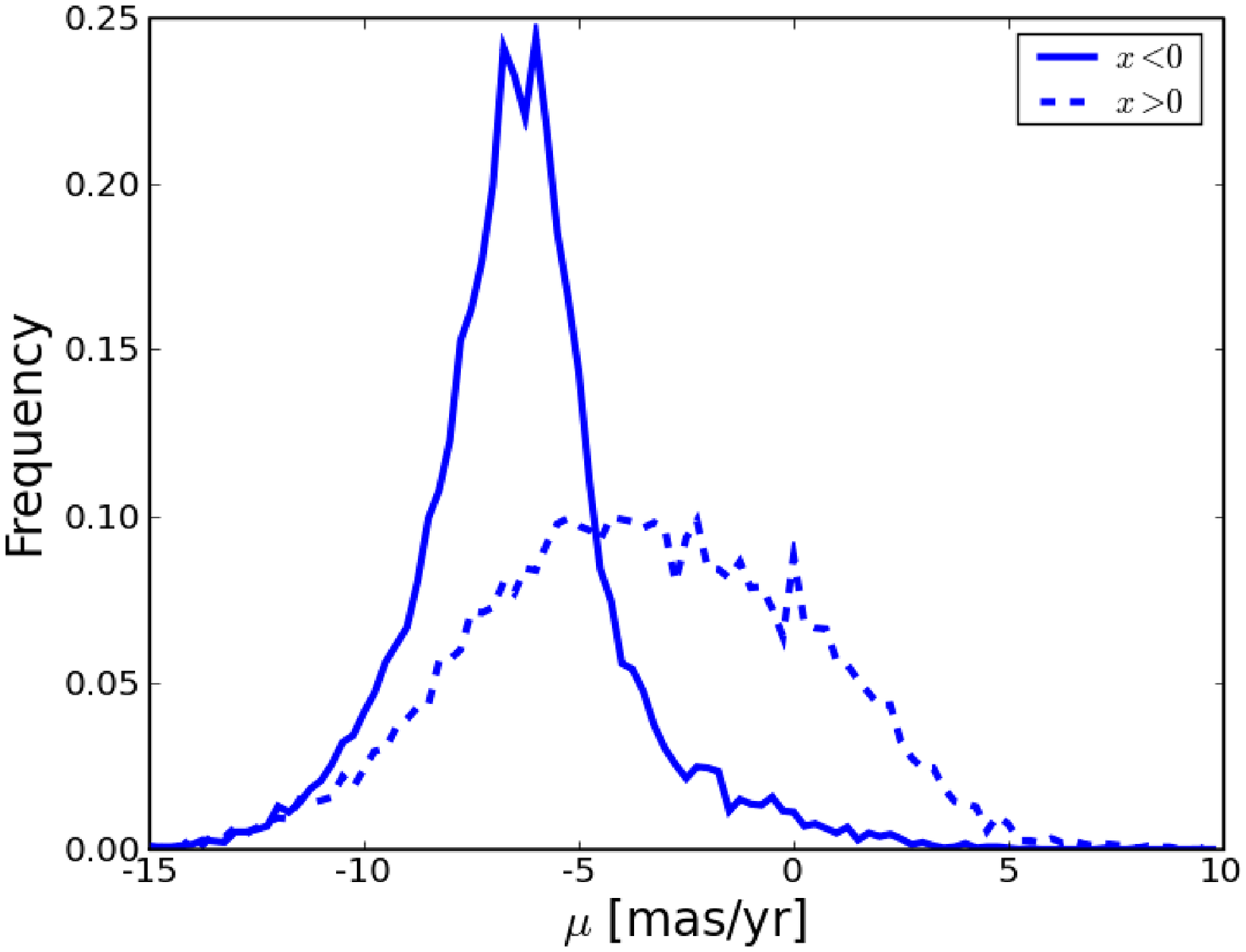}
   \includegraphics[width=6.08cm]{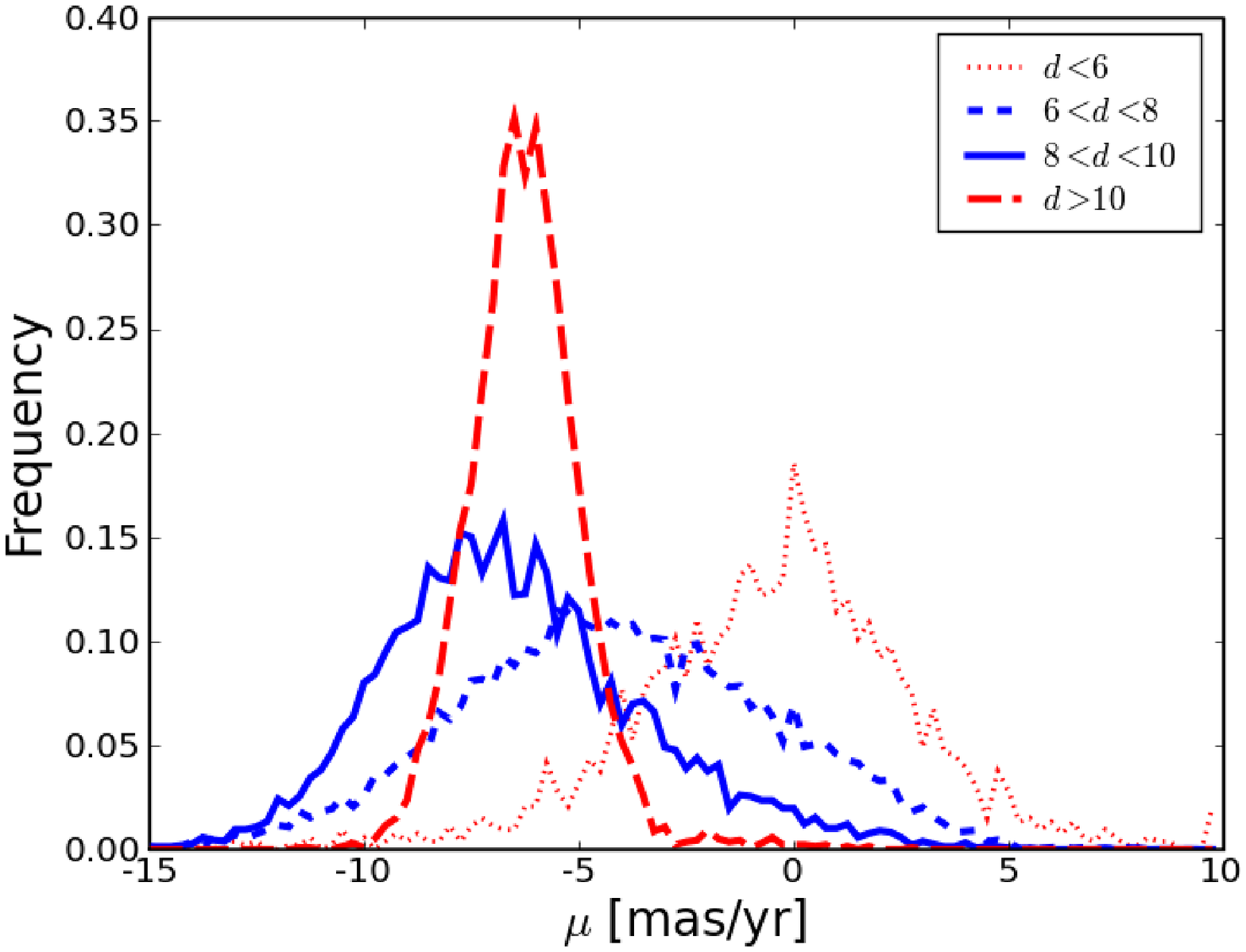}
   \caption{\small Distributions of radial velocities (left) and
     proper motions (middle) obtained by selecting stars
     with $0<l<20^{\circ}$ and positive/negative $x$ in the barred
     $N$-body model {\tt m08t3200}. Right panel: proper-motion
     distributions obtained by selecting stars with $0<l<20^{\circ}$
     and different distance $d$ from the observer.}
   \label{Fig9}%
\end{figure*} 

The differential velocity field ${\bf v}$ defined in
Eq.~(\ref{DiffVelField}) is now obtained by averaging the particle
velocities ${\bf V}_i$ located at different $z$ positions to smooth
out irregularities due to granularity, and by subtracting the averaged
velocity vector $\overline{\mathbf{V}}(\mathbf{x_0})\>$ of the
particles in a volume of 1~kpc$^3$ centered at the position of the
observer $\mathbf{x_0}$.  The radial velocity field and the tangential
velocity field are shown in Fig.~\ref{Fig7} (top and bottom panels,
respectively) and should be compared with the fields shown in
Fig.~\ref{Fig2} for the analytic axisymmetric model.  From these
panels it can be seen that, apart from noise in the boundary cells due
to the finite number of particles, the overall structure of the
contour maps is in agreement with those corresponding to the analytic
model. The presence of the bar does not introduce asymmetries changing
the overall results.  In order to quantify the effect of the bar, we
have constructed a symmetric system by randomising the particle polar
angles. Thus, we could evaluate that the relative error in both radial
and tangential velocity fields due to the presence of the bar is less
than 15\%.

The proper motion field (averaged on $z$) and the proper motion
$\mu_0$ (averaged on $z$ and $|y|<1$~kpc) are shown in Fig.~\ref{Fig8}
(left and right panels, respectively).  Also in this case the overall
structure of the contour map agrees with the analytic model map
(compare with Figs.~\ref{Fig3}) and the behavior of $\mu_0$ is analog
to that shown in Fig.~\ref{Fig4} (right panel) apart from numerical
oscillations due to the presence of the singular point at the
observer's position. The relative error between the barred model and
the one obtained by randomising angles is 6\%.  As in the axisymmetric
models, in Fux's barred model a region of particles with
large-amplitude proper motion exists, located behind the Galactic
center.  In this region, the proper motion is of the order of $\mu
\sim -7~\rm{mas \,yr}^{-1}$ for particles at distances to the observer
of 10~kpc. The same order of magnitude holds in regions slightly above
and below the Galactic plane, which can be reached by direct
measurements. This order of magnitude for the proper motion is well
within the astrometric accuracy of the GAIA survey, which will be
12-25~$\mu$as at the $15^{\rm th}$ magnitude and $100-300\,\mu$as at
the $20^{\rm th}$ magnitude (see e.g. Brown~\cite{brown}).

We compare now the predictions of this $N$-body model with some
observational results in order to understand how the parameters of the
Galactic bar can be constrained by measurements of radial velocities
and proper motions. Mao \& Paczy\'nski~(\cite{mao}) used two simple
analytic models to show that samples of stars in the far and near
sides of the bar display differences in both radial and tangential
streaming motions. In that paper, it was suggested to use red clump
giants to constrain the kinematics of stars in the Galactic bar. Red
clump giants are bulge stars which occupy a distinct region in the
color-magnitude diagram. They have a well-defined peak in their
observed luminosity function, which can be used to select stars at the
bright slope of the peak, and thus closer to us, from those at the
faint slope which are more distant.  Following this suggestion, data
of the Optical Gravitational Lensing Experiment II (OGLE-II) in the
Baade's window were used to derive the difference in the average
proper motion between the bright and faint samples of red clump giants
(Sumi et al.~\cite{sumi}) and the result $\Delta\langle\mu\rangle \sim
1.5~\rm{mas \,yr}^{-1}$ is consistent with the value estimated in Mao
\& Paczy\'nski~(\cite{mao}) who assumed a streaming motion of the bar
of $v\sim 100~\rm km\, s^{-1}$ in the same direction as the solar
rotation.  

More recently, different methods were used in order to select bulge
samples. Vieira et al.~(\cite{vieira}) measured proper motions inside
Plaut's window with an accuracy of $1\,\rm{mas \,yr}^{-1}$ using
ground-based data. They selected a bulge sample at the mean distance
of 6.37~kpc in the red giant branch by cross-referencing with the Two
Micron All Sky Survey catalog. Clarkson et al.~(\cite{clarkson})
reported on the use of the Advanced Camera for Surveys WFC on the
Hubble Space Telescope to extract proper motions in the Sagittarius
window with an accuracy of $0.3\,\rm{mas \,yr}^{-1}$.  They selected
kinematically a bulge sample by introducing a cutoff on longitudinal
proper motions and on proper-motion measurement errors, as suggested by
Kuijken \& Rich~(\cite{kuijken}). An interesting result of such
studies is that longitudinal proper motions can clearly separate near
and far samples of stars.

We use Fux's accurate model in order to evaluate the effect of
distance on radial-velocity and proper-motion measurements.  We
consider almost 24'000 particles with Galactic longitude $0< l <
20^{\circ}$ and we compare their distribution in the near region
$x>0$ with that in the far region $x<0$, averaged on $|z|<0.5$~kpc.
The distributions of radial velocities $v_r$ and proper motions $\mu$
are shown in the left and middle panels of Fig.~\ref{Fig9},
respectively.  From the left panel, it can be seen that the far sample
in $x<0$ is slightly shifted toward more negative radial velocities,
in qualitatively agreement with Mao \& Paczy\'nski~(\cite{mao}).  The
shift between the two samples is $\Delta\langle v_r\rangle\sim 5.4~\rm
km\, s^{-1}$. Longitudinal proper motions show much clearer
association with distance than radial velocities.  The middle panel in
Fig.~\ref{Fig9} shows that the distribution of far-sample proper
motions is shifted toward more negative values with respect to that of
the near sample, with a shift of $\Delta\langle\mu\rangle \sim
2.8~\rm{mas \,yr}^{-1}$. Moreover, the dispersion of the far sample is
60\% smaller than that of the near sample.

In order to discuss the effects of contamination of bulge samples by
disk stars we select stars in Fux's model with different mean
distances from the observer.  Since the corotation radius is $R_L =
4.8$~kpc and axis ratio is $b/a =0.5$ in Fux's model, samples in $0< l
< 20^{\circ}$ and $6<d<10$ (in kpc length unit) consist mainly of
bulge stars, while those in $d<6$ or $d>10$ have large contamination
by disk stars.  In the right panel of Fig.~\ref{Fig9} we show the
proper-motion distributions obtained by selecting stars within these
regions. By comparing the right and middle panels of Fig.~\ref{Fig9},
we see that the effect of the contamination from distant ($d>10$) and
nearby ($d<6$) disk stars is to move the mean value of the bulge
samples toward more positive values. Distant disk stars also have the
effect of reducing the dispersion of bulge proper-motion distribution,
as discussed in Vieira et al.~(\cite{vieira}). Knowledge of the global
distribution of proper motions as shown in Figs.~\ref{Fig3} and
\ref{Fig8} help us to understand and to evaluate these effects of
contamination.

\section{Conclusions}

Starting from the analysis of simple axisymmetric models of the Milky
Way, we have studied the differential velocity field, and in
particular the proper motion field, providing a global description of
observables.  Interestingly, there is a region behind the Galactic
center extending to several kpc further out where the proper motions
induced by differential rotation have the largest amplitudes in all
the Milky Way, an apparently little known basic feature (with the
exception of Binney\cite{binney}) with direct observational
consequences.  Here we clarify on a global scale the effects related
to first order to the average circular rotation and to second order to
a bar.

We have checked that these high proper motions are also found in
self-consistent barred configurations, such as the Milky Way $N$-body
model of Fux~(\cite{fux}).  The proper motions at distances of 10\,kpc
to the observer are of the order of 7\,$\rm{mas \,yr}^{-1}$ and thus
are well within the accuracy of the forthcoming surveys such as GAIA.
It is crucial to be able to separate the sources according to
distance, at least to separate the sources closer to or beyond the
Galactic center.

Provided that the position of the sources, such as stars or OH masers,
are intense enough to be measured with respect to an absolute frame of
reference, as well as their radial velocities, the global velocity
field of the Milky Way including non-circular motions should become
accessible.  This will be a first order information for constraining
the strength of the bar and the mass distribution in the bulge, the
keystone of the whole Galaxy.

\begin{acknowledgements}
  We thank Roger Fux for providing us the simulation data and Laurent
  Eyer for useful discussions.  We thank James Binney for constructive
  comments. This work has been supported by the Swiss National Science
  Foundation.
\end{acknowledgements}

\end{document}